# Ligand exchange in gold-coated FePt nanoparticles

P. de la Presa[*], T. Rueda[*], M. P. Morales[†] and A. Hernando[*]

[*]Instituto de Magnetismo Aplicado, UCM-ADIF-CSIC P.O. Box 155, 28260 Las Rozas, Madrid, Spain Email: see pdelapresa@adif.es [†]Inst. Ciencia Mat. Madrid, CSIC Sor Juana Ines de la Cruz 3, 28049 Madrid, Spain

*Abstract*—In this work we present the magnetic properties of gold coated FePt nanoparticles and the study of stable aqueous dispersions of FePt@Au and FePt synthesized after ligand exchange with mercaptoundecanoic acid. The particle size determined from TEM micrographs goes from 4 nm for the uncoated nanoparticles to a maximum of 10 nm for the gold coated ones indicating that the thickness of the shell ranges from 1 to 3 nm. The magnetic characterization consists in hysteresis cycles at 10 and 300 K. The results show that, at low field and room temperature, the magnetic behavior of uncoated and coated nanoparticles are surprisingly quite similar. Since the gold coated nanoparticles keep the magnetic properties of FePt and the presence of gold improves the functionalization of nanoparticles, the system is suitable for biological application. Mercaptoundecanoic ligand transfer was used to render water stable nanoparticles in a wide pH range. Transmission electron microscopy and dynamic light scattering results show the nanoparticles slightly agglomerate after ligand exchange. Fourier transform infrared spectroscopy results suggest that thiol bind to the gold atoms of the surface.

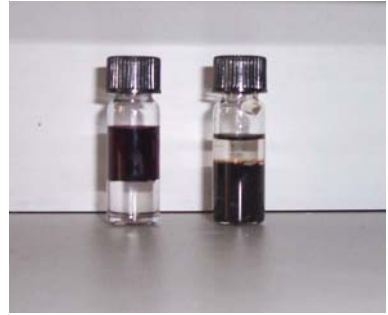

Fig. 1. (a) top: FePt@Au dispersed in hexane, bottom: water. (b) top: hexane, bottom: FePt@Au dispersed in water

## I. Introduction

The useful magnetic properties of FePt nanoparticles (NPs) arouse the interest of this material for biological applications[1], [2]. However, the synthetic procedure to obtain high quality NPs involves organic phase reactions. The NPs resulting from these procedures are stable in nonpolar solvents (such as hexane) and capped with nonpolar endgroups on their surface. The capping molecules (also called ligands) are typically long-chain alkanes with polar groups that bind to the nanoparticles' surface. To obtain a biocompatible and water dispersible compound the ligands must be exchanged by polar endgroups. In the case of FePt NPs, the difficulties arise from the fact that both, Fe and Pt atoms, have quite different affinities to bind organic molecules. Some efforts have been done to found a proper ligand, as for example mercaptoundecanoic acid (MUA), that can bind both Fe and Pt simultaneously[3].

Recently, we have improved the functionalization of FePt by coating the NPs with gold[4]; however, the coating procedure requires also an organic phase process resulting in FePt@Au capped by oleic acid and oleylamine and dispersed in hexane.
In this work we investigate the magnetic response of FePt and FePt@Au capped by oleic acid and oleilamine in order to study the influence of the gold coating on the magnetic properties of FePt, we find that at room temperature and low field the magnetic behavior is quite similar. Since MUA successfully displace oleic acid and oleylamine ligands in FePt and gold-coated Fe oxide NPs (Refs. [3], [7]), we choose this ligand to investigate the ligand-exchange in the gold-coated NPs. Contrary to MUA-exchanged FePt, in which the results suggest the binding of the carboxylate end to the iron atoms and the mercapto end to the platinum atoms, in the gold coated FePt NPs Fourier transform infrared spectroscopy (FTIR) suggests that thiol bind to the gold atoms of the surface. The transmission electron microscopy (TEM) and dynamic light scattering (DLS) results indicate that the ligand-exchanged NPs, both uncoated and coated, slightly agglomerate.

## II. Sample preparation

The synthesis of the FePt and FePt@Au Nps is described elsewhere[4], [8]. In order to make these NPs biocompatible, they must be transformed in water soluble NPs by changing the surfactant ligands. To perform the ligand exchange 1 g of MUA is dissolved in 5 ml chloroform, 10 mg of FePt@Au NPs dispersed in 5 ml chloroform is added to the first dispersion and a mild stirring is performed during 1 day; then, the NPs are precipitated with water by centrifuging. In a similar way, the oleic acid and oleilamine have been exchanged by MUA in FePt NPs. Figure 1 shows that the ligand exchanged NPs are successfully dispersed in water. In the Fig. 1(a), the FePt@Au NPs dispersed in hexane are in the top, whereas water is at the bottom. In the Fig. 1(b) the ligand exchanged FePt@Au NPs dispersed in water stay at the bottom of the bottle and hexane at the top.

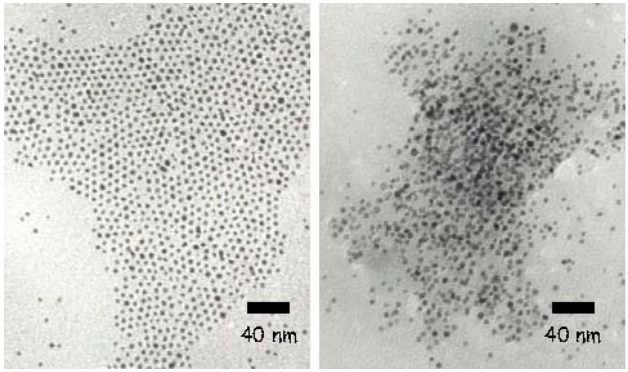
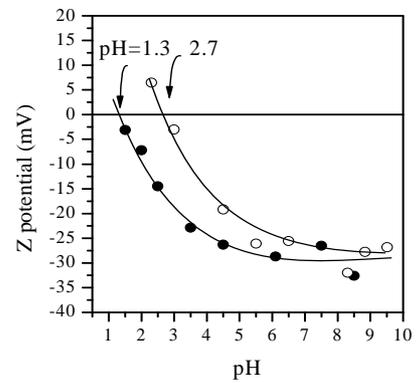

Fig. 2. Left: Oleic acid and oleylamine FePt@Au NPs dispersed in hexane. Right: MUA ligand exchanged FePt@Au NPs dispersed in water.

Fig. 3. Z potential of the MUA ligand-exchanged FePt (full circle) and FePt@Au (open circle) NPs.

TABLE I HYDRODYNAMIC RADIUS

| NPs | FePt | FePt@Au | MUA -FePt | MUA -FePt@Au |
|---|---|---|---|---|
| Dh (nm) | 5(1) | 7-12(1) | 150 -200(50) | 170 -240(80) |

## III. SAMPLE CHARACTERIZATION

### A. TEM

The particle size was determined from TEM micrographs in a 200 keV JEOL-2000 FXII microscope. For the observation of the sample in the microscope, a drop of the dispersed suspension of particles was placed onto a copper grid covered by a carbon film. The mean particle size was calculated by counting more than 100 particles. The diameter of the NPs increases from 4 nm for the uncoated NPs to a maximum of 10 nm for the gold coated ones indicating that the thickness of the shell ranges from 1 to 3 nm. Figure 2 shows the TEM image of the oleic acid and oleylamine capped FePt@Au (left) and the MUA ligand-exchanged NPs (right). Energy dispersive x-ray (EDX) analysis yields a mean composition of $Fe_{55}Pt_{45}$ and about 60%at. of gold in the gold coated NPs.

### B. Dynamic light scattering

We performed dynamic light scattering (DLS) measurements in a ZETASIZER NANO-ZS device (Malvern Instruments) to determine the hydrodynamic radius of the uncoated, gold coated and the ligand-exchanged NPs. Hydrodynamic size was measured at pH 7. Table I shows the hydrodynamic radius measured in FePt, FePt@Au capped with oleic acid and oleylamine and MUA-exchanged FePt and FePt@Au.

Electrophoretic mobility was measured as a function of pH at 25 °C, using $10^{-2}$ M $KNO_3$ as electrolyte and $HNO_3$ and KOH to change the pH. Electrophoretic mobility was measured in the same apparatus in a special Zeta Potential cuvette. Data were transformed to zeta potential values, which are related to the surface charge density and depend on the composition, crystalline form, size and surface characteristics[5]. The surface charge for both samples at pH 7 is higher than 25 mV, which guarantees its stability for a long time[6]. Z potential (mV) 20 15 5 0

### C. FTIR spectroscopy

A Nicolet 20SXC FT-IR was used to collect the vibrational spectra of the as-made and MUA exchanged NPs. A FTIR spectrum of FePt@Au, was collected after removal of the excess surfactants by washing in ethanol and separating the NPs from solution by centrifugation, the powder sample was mixed with KBr powder and then pressed into pellets. Figure 4 shows FTIR spectra of FePt@Au capped by oleic acid and oleylamine and MUA-exchanged NPs.

### D. Magnetic characterization

The FePt and FePt@Au coated by oleic acid and oleylamine have been magnetically characterized by mean of a Quantum Design SQUID magnetometer. The magnetic characterization consists in hysteresis cycles at 10 and 300 K and zero fieldcooled (ZFC) and fiel-cooled (FC) curves at 100 Oe. The ZFC-FC curves show that both, FePt and FePt@Au NPs, are superparamagnetic with blocking temperatures 25 and 12 K, respectively[4]. Since the NPs are superparamagnetic, there is not magnetic saturation and the magnetic curves are normalized for each temperature to the magnetic value at higher applied field of the uncoated NPs (see Fig. 5).

## IV. RESULTS AND DISCUSSION

The normalization of the hysteresis curve to the magnetic value of FePt Nps at 5 T allows to determine the relative reduction of the magnetization in FePt@Au due to the presence of gold. The magnetic response of the uncoated and goldcoated NPs differs considerably at high and low fields. The magnetic moment at H = 5 T at 10 and 300 K for FePt@Au is reduced to 28 and 23%, respectively. The amount of gold in the gold-coated NPs is about 70%wt., therefore, the reduction of the magnetization at high field can be mainly attributed to the presence of diamagnetic gold atoms. However, at lower fields, the behavior is quite different. At low temperature, the coercive field decreases from 1300 to 350 Oe in the gold coated NPs (inset bottommost Fig. 5), this reduction together with the reduction of the blocking temperature can be related to the reduction of the magnetic anisotropy due to the presence of diamagnetic gold atoms at the surface of the magnetic core FePt. On the other hand, at room temperature, the magnetic behavior of the uncoated and gold-coated NPs is quite similar at low fields (H < 2 kOe, see inset topmost Fig. 5). This result makes the gold coated NPs interesting for biological applications in which low applied fields are required, as for example, magneto fluid hyperthermia (MFH). The applied fields in MFH treatment are much lower than 1 kOe (Ref. [9]), since FePt and FePt@Au have similar magnetic behavior at room temperature and low fields, the gold coating improves the functionalization of the FePt keeping the magnetic properties, making the gold coating of FePt an interesting root for MFH applications.

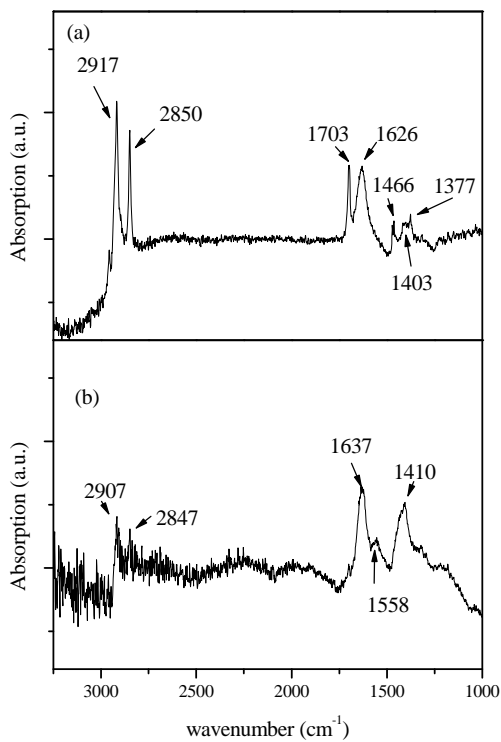

Fig. 4. FTIR spectra of (a) oleic acid and oleylamine FePt@Au and (b) MUA-exchanged FePt@Au.

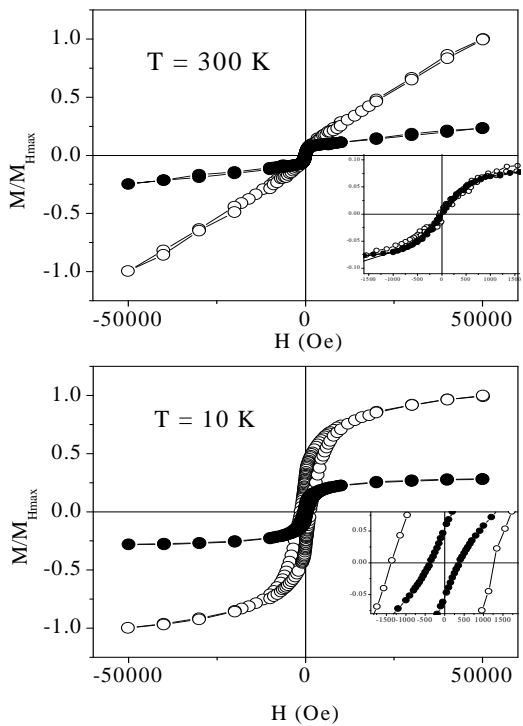

Fig. 5. Hysteresis cycles of FePt (open circle) and FePt@Au (full circle) NPs. The hysteresis curves are normalized to the magnetization of FePt Nps for each temperature.

As can be seen in Fig 2, TEM images of both uncoated and coated FePt NPs, the NPs capped by oleic acid and oleylamine tend to self assemble. By comparing TEM images of the oleic acid and oleylamine FePt@Au and MUAexchanged NPs, it can be observed the NPs dispersed in hexane tend to self-assemble maintaining a regular distance between them, whereas in the MUA exchanged NPs there is no sign of pattern, indicating that FePt and the FePt@Au slightly agglomerate after the ligand-exchange.

From DLS measurements it was determined that the oleic acid and oleylamine FePt have hydrodynamic diameter (Dh) of about 4-5 nm, whereas Dh ranges from 7 to 12 nm for gold coated NPs. Regarding the length of the oleic acid and oleylamine ligands (about 1 nm), these values are in agreement with well dispersed NPs with an organic layer about 1 nm in size surrounding the surface. On the other hand, the ligandexchanged NPs have a much greater Dh (as can be seen in Table I) indicating that the MUA-exchanged NPs, both the FePt and the FePt@Au, slightly agglomerate, in agreement with TEM results. These agglomerates are about 200 nm in size.

The MUA-exchanged NPs are negatively charged for pH > 1.7 (FePt) and pH > (2.7 (FePt@Au), as can be seen in Fig. 3. However, the more alkaline is the water, the more negative charges are at the surfaces, the more stable are the NPs in water. This in agreement with the report on MUA-ligand exchanged FePt by Bagaria et al. [3], the authors reported that the NPs are stable in highly alkaline water. It should be emphasized that the colloidal suspensions prepared in this work are highly stable at pH 7 thanks to its high surface charge, which is an important requirement for biomedical applications

The FTIR spectrum of the as-made FePt@Au NPs has peaks around 3000 cm$^{-1}$ which are characteristic for both, oleic acid and oleylamine, and are assigned to the symmetric and asymmetric -CH stretches[10]. The peak at 1626 cm$^{-1}$ is related to the ν(C=C), also common to both organic molecules. In Fig. 4(a), the peak at 1703 cm$^{-1}$ can be assigned to the stretch mode ν(C=O), indicating that oleic acid is bonded to the as prepared NPs. In the FTIR spectra, the scissoring mode associated to molecularly bonded oleylamine normally observed at 1590 cm$^{-1}$ is not possible to discriminate.

The change in the intensity ratio of the -CH stretches around 3000 cm$^{-1}$ to the peaks around 1500 cm$^{-1}$ in the ligandexchanged NPs compared to the as-synthesized ones already indicates that the surface chemistry has changed. The peak at 1637 cm$^{-1}$ is the ν(C=C) mode. Furthermore, the 1703 cm$^{-1}$ peak related to monodendate vanishes in the ligand exchanged NPs, whereas the 1558 cm$^{-1}$ peaks agrees well with the oleate ν(COO$^-$) mode in standard sodium oleate FTIR spectrum[11], this probably suggests that the carboxylate end of MUA does not bind to the gold surface. Further investigations are underway in order to clarify this point.

## CONCLUSIONS

The gold coated FePt NPs capped by non-polar end groups (as oleic acid and oleylamine) and dispersed in hexane have been successfully ligand-exchanged by MUA, which makes the NPs stable in water. The FTIR results suggest that the thiol end of MUA binds to gold atoms instead of the carboxylate end. The magnetic properties of gold coated FePt at low field and room temperature are similar to the "as-synthesized" FePt. Therefore, the MUA-exchanged gold coated NPs can be of interest for hyperthermia applications, in which water dispersible NPs are required and fields lower than 1 kOe are applied. Gold coating not only provide a versatic cover for functionalization with different biomolecules but also would allow heating and therefore transforming the FePt nanoparticles in its high anisotropy phase without sintering, leading to a promising material for hyperthermia treatments.

## ACKNOWLEDGMENT

Financial support from the Comunidad Autónoma de Madrid under Project No. S-0505/MAT/0194 and Ministerio de Educación y Ciencia Grants NAN2004-09125-C07-05 and MAT2005-06119 are acknowledged.